\documentclass[useAMS,usenatbib]{mn2e}

\usepackage{amsmath,graphicx}
\usepackage{tabularx, booktabs}
\usepackage{comment}

\newcommand{\aap}{A\&A}
\newcommand{\aj}{AJ}
\newcommand{\apj}{ApJ}
\newcommand{\apjl}{ApJL}
\newcommand{\apjs}{ApJS}
\newcommand{\araa}{ARA\&A}
\newcommand{\mnras}{MNRAS}
\newcommand{\na}{New Astron.}
\newcommand{\nar}{New Astron. Rev.}

\newcommand{\zap}{Z. Astrophys.}

\newcommand{\kms}{\text{km~s}^{-1}}

\newcommand{\cmmm}{\text{cm}^{-3}}

\newcommand{\msunyr}{\text{M}_\odot\,\text{yr}^{-1}}

\newcommand{\HH}{\text{H}_2}          
\newcommand{\HM}{{\rm H}^{-}}     
\newcommand{\HP}{{\rm H}^{+}}     
\newcommand{\HII}{H{\sc ii}~}    
\newcommand{\e}{{\rm e}^{-}}     

\newcommand{\ob}{\Omega_{\rm b}}
\newcommand{\ol}{\Omega_\Lambda}
\newcommand{\om}{\Omega_{\rm m}}

\newcommand{\nh}{n_{\rm H}}

\newcommand{\msun}{\text{M}_\odot}
\newcommand{\rsun}{\text{R}_\odot}

\newcommand{\mdm}{M_{\rm dm}}

\newcommand{\rvir}{R_{\rm vir}}

\newcommand{\mvir}{M_{\rm vir}}

\newcommand{\mbe}{M_{\rm BE}}
\newcommand{\menc}{M_{\rm enc}}

\newcommand{\cs}{c_{\rm s}}

\newcommand{\geff}{\gamma_{\rm eff}}
\newcommand{\vrot}{v_{\rm rot}}
\newcommand{\vrad}{v_{\rm rad}}
\newcommand{\vkep}{v_{\rm kep}}
\newcommand{\mstardot}{\dot{M}_\star}
\newcommand{\mstar}{M_\star}
\newcommand{\rstar}{R_\star}

\title[Massive protostars in atomic cooling haloes]{Formation of massive protostars in atomic cooling haloes}

\author[Becerra et al.]{\parbox{17.5cm}{Fernando Becerra$^1$\thanks{E-mail: fbecerra@cfa.harvard.edu}, Thomas H. Greif$^1$, Volker Springel$^{2,3}$, Lars E. Hernquist$^1$}
\\$^1$ Harvard-Smithsonian Center for Astrophysics, 60 Garden Street, Cambridge, MA 02138, USA
\\$^2$ Heidelberg Institute for Theoretical Studies, Schloss-Wolfsbrunnenweg 35, D-69118 Heidelberg, Germany
\\$^3$ Zentrum f\"ur Astronomie der Universit\"at Heidelberg, ARI, M\"onchhofstr. 12-14, D-69120 Heidelberg, Germany}

\begin{document}

\maketitle
\topmargin-1cm

\begin{abstract}
We present the highest-resolution three-dimensional simulation to date of the collapse of an atomic cooling halo in the early Universe. We use the moving-mesh code {\sc arepo} with the primordial chemistry module introduced in \citet{Greif_2014}, which evolves the chemical and thermal rate equations for over more than 20 orders of magnitude in density. Molecular hydrogen cooling is suppressed by a strong Lyman-Werner background, which facilitates the near-isothermal collapse of the gas at a temperature of about $10^4\,$K. Once the central gas cloud becomes optically thick to continuum emission, it settles into a Keplerian disc around the primary protostar. The initial mass of the protostar is about $0.1\,\msun$, which is an order of magnitude higher than in minihaloes that cool via molecular hydrogen. The high accretion rate and efficient cooling of the gas catalyse the fragmentation of the disc into a small protostellar system with $5-10$ members. After about $12\,$yr, strong gravitational interactions disrupt the disc and temporarily eject the primary protostar from the centre of the cloud. By the end of the simulation, a secondary clump has collapsed at a distance of $\simeq 150\,$au from the primary clump. If this clump undergoes a similar evolution as the first, the central gas cloud may evolve into a wide binary system. High accretion rates of both the primary and secondary clumps suggest that fragmentation is not a significant barrier for forming at least one massive black hole seed.
\end{abstract}


\section{Introduction}
\label{sec:intro}

Black holes (BHs) are a key ingredient in the formation and evolution of galaxies. In the local Universe, the stellar velocity dispersion in galaxy bulges is correlated with the mass of the BH at their centre \citep{Ferrarese_00,Gebhardt_00}. BHs also power luminous quasars by accreting gas from their host galaxies. Recent observations suggest that quasars powered by BHs with masses $\ga 10^9\,\msun$ were already present when the Universe was less than one billion year old \citep{Fan_2003, Fan_2006}. These supermassive black holes most likely grew from smaller seed BHs that formed earlier, but the origin of these seeds remains unclear \citep{Haiman_2006, Haiman_2009, Greene_2012, Volonteri_2012, Volonteri_Bellovary_2012}. One possible candidate are the remnants of massive Population~III stars \citep{Madau_2001, Li_2007, Johnson_2012}, or the direct collapse of primordial gas in haloes with virial temperatures $T_{\rm vir}\ga 10^4\,$K, so-called atomic cooling haloes \citep{Bromm_2003, Bromm_Yoshida_2011}. In the former case, the seeds have initial masses of the order of $100\,\msun$, and grow at or above the Eddington limit for the remaining $\simeq 500\,$Myr between seed formation and $z\simeq 6$. However, numerical simulations have shown that accretion on to early BHs is inefficient, due to the low density of the gas surrounding the BH remnant, which is caused by photoionization heating from the progenitor star \citep{Johnson_2007, Alvarez_2009}. Accretion rates are thus not high enough to allow efficient growth of the seed, which poses a serious complication for the Population~III stellar remnant scenario.

In the direct collapse scenario, haloes with virial temperatures $\ga 10^4$ K may host seed BHs that are substantially more massive. A prerequisite is that the accretion rate on to the central object is high enough that radiative feedback does not severely impede the accretion flow \citep{Johnson_2011,Johnson_2012, Hosokawa_2012, Hosokawa_2013}. In this case, a supermassive star or `quasi-star' forms, which may collapse into a BH of mass $\sim 10^5-10^6\,\msun$ \citep{Heger_2003, Begelman_2006, Begelman_2008, Begelman_2010, Volonteri_Begelman_2010, Montero_2012, Volonteri_2012, Inayoshi_2013, Schleicher_2013, Chen_2014}. Since the accretion rate in a Jeans-unstable cloud scales as ${\dot M}\propto T^{3/2}$, molecular hydrogen cooling must be suppressed until the virial temperature of the halo is high enough that Ly$\alpha$ cooling becomes important. This may be achieved by a Lyman-Werner (LW) radiation background \citep{Omukai_2001, Bromm_2003, Volonteri_2005, Spaans_2006, Schleicher_2010, Johnson_2013}. Simple one-zone models have found that the critical flux is of the order of $J_{21, {\rm crit}}=10^5$ in units of $J_{21} = 10^{-21}\,{\rm erg}\,{\rm s}^{-1}\,{\rm cm}^{-2}\,{\rm Hz}^{-1}\,{\rm sr}^{-1}$ for a blackbody spectrum with $10^5\,$K \citep{Omukai_2001}. For Population~I/II stars, recent studies have found that the critical flux may be somewhat lower \citep{Shang_2010, Wolcott-Green_2012, Van_Borm_2013, Agarwal_2014, Latif_2014a, Latif_2014c, Regan_2014b, Sugimura_2014}. Even though the LW flux on cosmological scales is well below this value, local star formation may raise the flux to supercritical levels \citep{Dijkstra_2008, Dijkstra_2014, Agarwal_2012, Agarwal_2014b, Visbal_2014}.

If the LW flux is high enough, the halo gas collapses nearly isothermally at $\simeq 10^4\,$K up to a density of $\nh\simeq 10^6\,\cmmm$, where the gas becomes optically thick to Ly$\alpha$ emission \citep{Omukai_2001}. At this point, continuum cooling via free-bound emission of H$^-$ takes over, and allows the gas to again contract nearly isothermally up to a density of $\nh\simeq 10^{16}\,\cmmm$. Once the continuum emission becomes trapped, the gas evolves nearly adiabatically and a protostar forms at the centre of the halo. During the initial collapse, the angular momentum is constantly redistributed by turbulence and bar-like instabilities, such that the cloud contracts nearly unhindered \citep{Oh_2002, Koushiappas_2004, Begelman_2006, Lodato_2006, Wise_2008, Begelman_2009, Choi_2013, Latif_2013a, Prieto_2013}.

The subsequent accretion phase was investigated by \citet{Regan_2009} and \citet{Latif_2013b, Latif_2013a}. They found that a Keplerian disc forms around the primary protostar, which becomes gravitationally unstable and fragments into a small system of protostars. The secondary protostars merge on a short time-scale and do not prevent the growth of the primary protostar. These studies employed a pressure floor beyond a certain refinement level, such that the maximum density was limited to $\nh\sim 10^6-10^9\,\cmmm$. The simulations of \citet{Regan_2014a} also displayed the formation of a disc-like object at the centre of the halo, which in some cases fragmented on a scale of $100\,$au. However, these simulations also suffered from limited resolution, and did not include the relevant $\HH$ cooling and chemistry. Recently, \citet{Inayoshi_2014} used the most detailed chemical and thermal model to date, but stopped the simulation once the primary protostar had formed. In addition, they did not use cosmological initial conditions. We here attempt to improve upon these studies by carrying out a simulation that starts from cosmological initial conditions and is not resolution-limited. We use a slightly less sophisticated chemical model as \citet{Inayoshi_2014}, but evolve the simulation well beyond the formation of the first protostar at the centre of the halo.

Our paper is organized as follows. In Section~\ref{sec:simulations}, we describe the simulation setup and the chemistry and cooling network. In Section~\ref{sec:results}, we analyse the simulation and discuss the collapse of the central gas cloud, the formation and fragmentation of the disc, the development of the protostellar system, and the collapse of a secondary clump towards the end of the simulation. Finally, in Section~\ref{sec:conclusion} we summarize and draw conclusions. All distances are quoted in proper units, unless noted otherwise.

\section{Simulations}
\label{sec:simulations}

We perform three-dimensional, cosmological hydrodynamical simulations to investigate the collapse of gas in atomic cooling haloes in which the formation of $\HH$ has been suppressed by a LW background. For this purpose we employ the moving-mesh code {\sc arepo} \citep{Springel_2010}. We also include the recently developed primordial chemistry and cooling network of \citet{Greif_2014}. In the following, we briefly describe the initialization of the simulations, the extraction procedure and refinement criteria used to achieve densities $\nh \ga 10^{21}~\cmmm$, and the chemistry and cooling network.

\subsection{Dark matter simulations}
\label{subsec:DMsimulations}

We first initialize a dark matter (DM)-only simulation at a redshift of $z = 99$ in a standard $\Lambda$ cold dark matter ($\Lambda$CDM) cosmology. We adopt cosmological parameters based on the \emph{Wilkinson Microwave Anisotropy Probe} results \citep{Komatsu_2009}. We use a matter density $\om = 1- \ol = 0.27$, baryon density $\ob = 0.046$, Hubble parameter $h = H_0/100~\kms~\text{Mpc}^{-1} = 0.7$ (where $H_0$ is the present Hubble expansion rate), spectral index $n_{\rm s} = 0.96$, and normalization $\sigma_8 = 0.81$. The simulation is initialized in a box of side length $2\,$Mpc (comoving) with a total of $512^3$ DM particles of mass $\simeq 2.2\times 10^3~\msun$. The gravitational softening length is set to $\simeq 195$ pc (comoving), which corresponds to 5\% of the initial mean inter-particle separation. We stop the simulation when the first halo with virial mass exceeding $10^8\,\msun$ collapses. This occurs at $z_{\rm coll} \simeq 12.4$, when the first halo reaches $\mvir \simeq 1.7 \times 10^8\,\msun$. At this point the halo has a virial radius of $\rvir \simeq 1.4\,$kpc and a spin parameter $\lambda \simeq 0.05$.

\subsection{Resimulations}
\label{subsec:resimulations}

The second step is to locate the target halo and flag it for further refinement. We select the particles belonging to that halo and a sufficiently large boundary region around it, and trace them back to their initial conditions. Once the particle locations have been determined, we reinitialize the simulation centred on the target halo. In order to acquire higher resolution we replace each DM particle by 64 less-massive DM particles and 64 mesh-generating points. The resolution is gradually decreased as the distance from the high-resolution region increases, replacing cells and DM particles by higher-mass particles outside the target region. The resimulation has lower resolution towards the edges of the box than the original DM-only simulation, but the accuracy of the gravitational tidal field around the target halo is preserved. The refined DM particle mass is given by $M_{\rm dm,ref} = (1 - \ob/\om)\mdm/64 \simeq 28~\msun$, and the gravitational softening length is set to $\simeq 49\,$pc (comoving). The refined mass of each cell is given by $M_{\rm gas,ref} = (\ob/\om)\mdm/64 \simeq 6\,\msun$.

We stop the resimulation once the first cell has exceeded a density of $\nh \simeq 10^9\,\cmmm$. We then proceed to extract the particles in the central $3\,$pc and reinitialize the simulation with reflective boundary conditions. Hence, the central region of the final output in the first resimulation becomes the initial condition for a second resimulation with a box size of $3\,$pc. Furthermore, at those densities the gas component is already well decoupled from the DM component \citep{Greif_2011}, so we discard the DM and keep only the gas particles. We evolve the second resimulation until it exceeds a density of $\nh \simeq 10^{19}\,\cmmm$, after which we conduct a second extraction similar in nature to the first, but cut out the central $5\times 10^{-3}\,$pc of the second resimulation. We then use this as the side length for the third resimulation. This approach has the risk that perturbations from the edges of the box might influence the central regions. However, we explicitly avoid this issue by assuring that the sound crossing time through the box is much longer than the free-fall time of the central high-density cloud.

\subsection{Refinement}
\label{subsec:refinement}

An essential refinement criterion that grid codes have to fulfill to resolve gravitational instability and avoid artificial fragmentation is the so-called Truelove criterion \citep{Truelove_1997}. This criterion states that the local Jeans length needs to be resolved by at least four cells, where the cell size is approximately given by $h = (3V/4\pi)^{1/3}$, and $V$ is the volume of the cell. In order to adequately resolve turbulence, recent studies using grid codes with a fixed mesh have found that the Jeans length must be resolved by at least 32 cells \citep{Federrath_2011, Turk_2012, Latif_2013a}. A disadvantage of using refinement based on the Jeans length is that shock-heated regions may be much less resolved than adjacent cold regions. In order to avoid this problem, we follow the refinement criterion proposed by \citet{Turk_2010}, who suggest using the minimum temperature of the gas to evaluate the Jeans length. We slightly modify this criterion by using $T_{\rm min} = 5000\,$K for cells with $T \le T_{\rm min}$, but the correct temperature for cells with $T>T_{\rm min}$. This ensures that the initial collapse phase is adequately resolved, while at high densities the resolution does not become excessively high and slows down the calculation. Below $\nh=10^{15}\,\cmmm$, we employ 64 cells per Jeans length, which is degraded to 8 cells above $\nh=10^{18}\,\cmmm$. We use a linear extrapolation between these densities. The maximum spatial resolution achieved with this refinement strategy is $\simeq 6.6 \times 10^{-4}\,$au. Next to the Jeans refinement, we refine a cell if its mass increases to more than twice its initial mass.

\subsection{Chemistry and cooling}
\label{subsec:chemistry}

A detailed description of the chemical and thermal model used here can be found in \citet{Greif_2014}. Here, we only briefly describe the most important reactions and cooling processes. The chemical network employs a non-equilibrium solver at low densities and an equilibrium solver at high densities for the species H, $\HH$, $\HM$, $\HP$, and $\e$. The transition from non-equilibrium to equilibrium H$_2$ chemistry occurs at $n_{\rm H_2, eq} = 10^{15}~\cmmm$, since three-body reactions depend on the cube of the density and would otherwise prohibitively decrease the time-step of the non-equilibrium solver. For densities above $n_{\rm H^+, eq} = 10^{18}~\cmmm$, the electron and $\HP$ abundances are also considered to be in equilibrium. The main reactions include the formation of $\HH$ via associative detachment as well as three-body reactions, the destruction of $\HH$ via collisions and photodissociation, and the formation and destruction of $\HP$ by collisional ionizations and recombinations.

\begin{figure*}
\begin{center}
\includegraphics[scale=0.68]{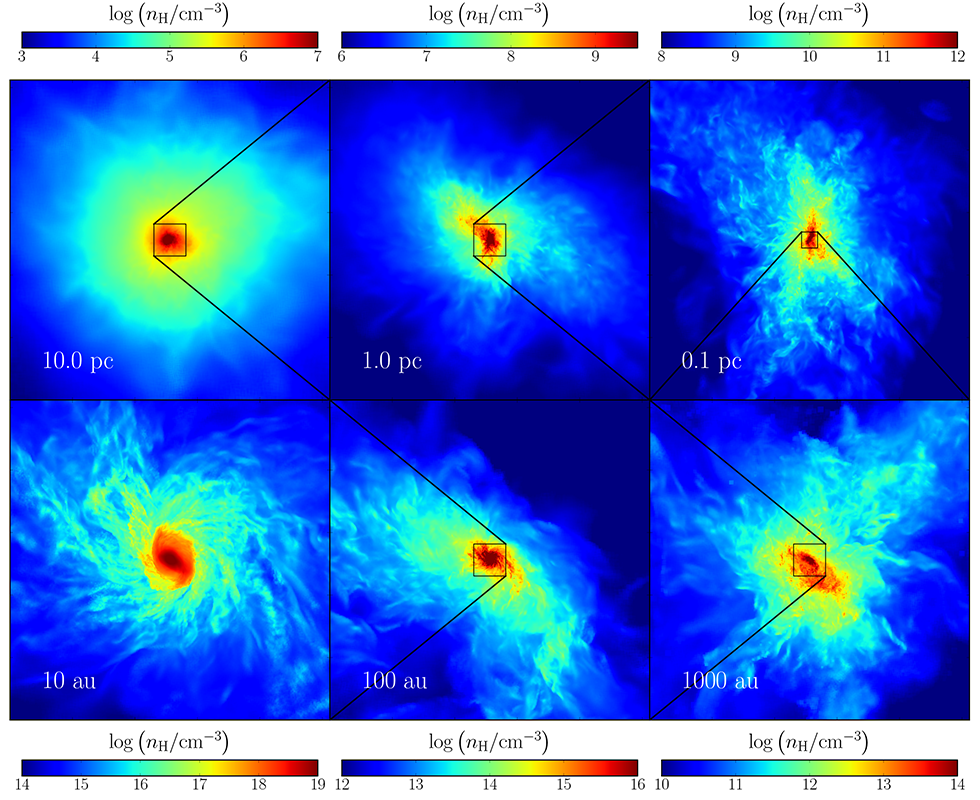}
\caption{Zoom-in on the gas cloud that forms at the centre of the atomic cooling halo. The number density of hydrogen nuclei is weighted with the square of the density along the line of sight, which is perpendicular to the plane of the disc. Clockwise from the top left, the width of the individual cubes are $10\,$pc, $1\,$pc, $0.1\,$pc, $1000\,$au, $100\,$au, and $10\,$au. The cloud has an irregular morphology that continues to change shape and orientation throughout the collapse. The filamentary structure indicates that turbulence is present on all scales.}
\label{fig:collapse}
\end{center}
\end{figure*}

The relevant cooling processes are $\HH$ line cooling, $\HH$ collision-induced emission, Ly$\alpha$ cooling, and inverse Compton cooling. $\HH$ cooling plays a substantial role up to $\nh\simeq 10^{15}~\cmmm$, where the gas becomes optically thick to the $\HH$ line emission, while collision-induced emission becomes important at $\nh\ga 10^{14}~\cmmm$ and provides the last radiative cooling channel \citep{Omukai_1998, Ripamonti_2004}. Although we include molecular hydrogen cooling, its effect does not become important during the evolution of the simulation due to the presence of a strong LW background that dissociates $\HH$ via the Solomon process \citep{Abel_1997}. Previous studies found that a strong LW flux with $J_{\rm 21}\ga 10^3$ is required to dissociate molecular hydrogen in the progenitors of an atomic cooling halo \citep{Omukai_2001, Johnson_2007, Dijkstra_2008, Latif_2013b, Wolcott-Green_2011}. Here, we assume a constant LW flux of $J_{\rm 21}=10^5$ for a blackbody spectrum with $T_{\rm rad}=10^5\,{\rm K}$, which is commonly used to estimate the spectra of Population III stars. In this case, the $\HM$ photodissociation rate is much smaller than the $\HH$ photodissociation rate \citep{Sugimura_2014}. We approximate the combined effects of Ly$\alpha$ cooling and continuum cooling by assuming that Ly$\alpha$ cooling remains optically thin up to densities $\nh\simeq 10^{16}\,\cmmm$. The cooling rate is exponentially suppressed at densities $\nh\simeq 10^{16}\,\cmmm$ to approximately reproduce the density-temperature relation found in \citet{Omukai_2001}. Due to this simplification, we may somewhat underestimate the true cooling rate.

\section{Results}
\label{sec:results}

\subsection{Collapse of central gas cloud}
\label{subsec:collapse}
        
A number of studies have discussed the properties of the collapse of primordial gas clouds in atomic cooling haloes \citep[e.g.,][]{Bromm_2003, Regan_2009, Choi_2013, Latif_2013a, Inayoshi_2014, Regan_2014a}. Here, we investigate the collapse of the gas over an unprecedented range in scale, as shown in Fig.~\ref{fig:collapse}. The six panels show a zoom-in on the central gas cloud, ranging from $10\,$pc down to scales of $10\,$au. The panel on the bottom-left side of the figure shows the primary protostar surrounded by an accretion disc. The cloud shows an irregular morphology and changes shape as it collapses. Its filamentary structure is indicative of turbulence, which is especially pronounced during the later stages of the collapse. On the largest scales, the cloud shows less substructure and is more spherically symmetric.
        
\begin{figure*}
\begin{center}
\includegraphics[scale=0.9]{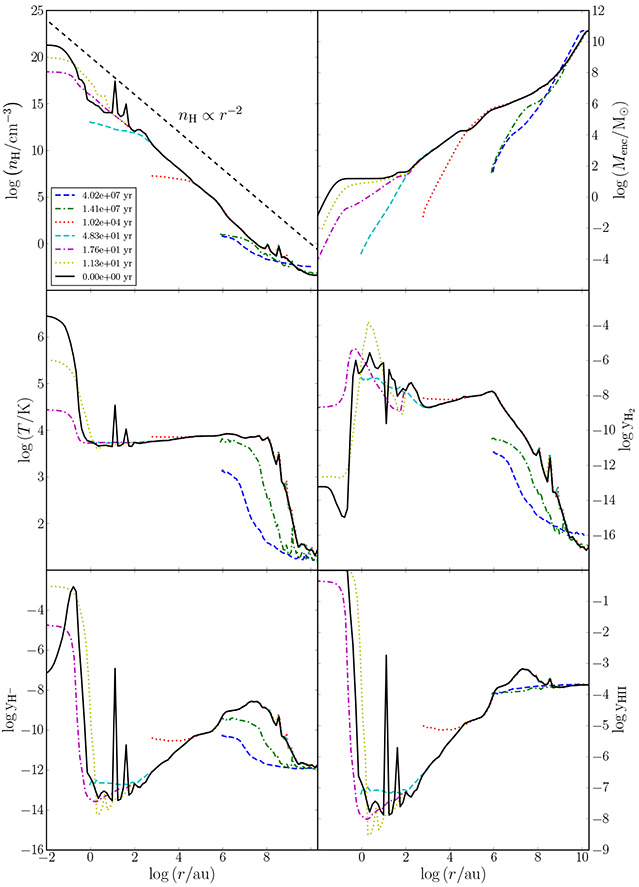}
\caption{Radial profiles for the mass-weighted number density of hydrogen nuclei, enclosed gas mass, temperature, and $\HH$, $\HM$, and \HII abundances between about $0.01\,$au and $\sim 100\,$kpc. The various line styles and colours denote different epochs of the evolution of the halo, labelled by their lookback time as measured from the end of the run according to the legend. The blue dashed line corresponds to redshift $\simeq 26$, the green dash-dotted line to $z \simeq 15$, the red dotted line to when the number density first exceeds $10^9\,\cmmm$, the cyan dashed line to $\simeq 9\times 10^3\,$yr after that, the purple dash-dotted line to when the number density first exceeds $10^{19}\,\cmmm$, the yellow dotted line to $\simeq 6\,$yr after the formation of first protostar, and the black solid line to the end of the simulation after approximately $18\,$yr. The halo follows several evolutionary stages from large to small scales: shock-heating to the virial temperature, onset of cooling, Jeans instability, isothermal contraction, formation of the primary protostar and disc, and fragmentation of the disc (see Section \ref{sec:results} for details).}
\label{fig:nh_enc_mass_temp_abH2}
\end{center}
\end{figure*}

\begin{figure*}
\begin{center}
\includegraphics[scale=0.61]{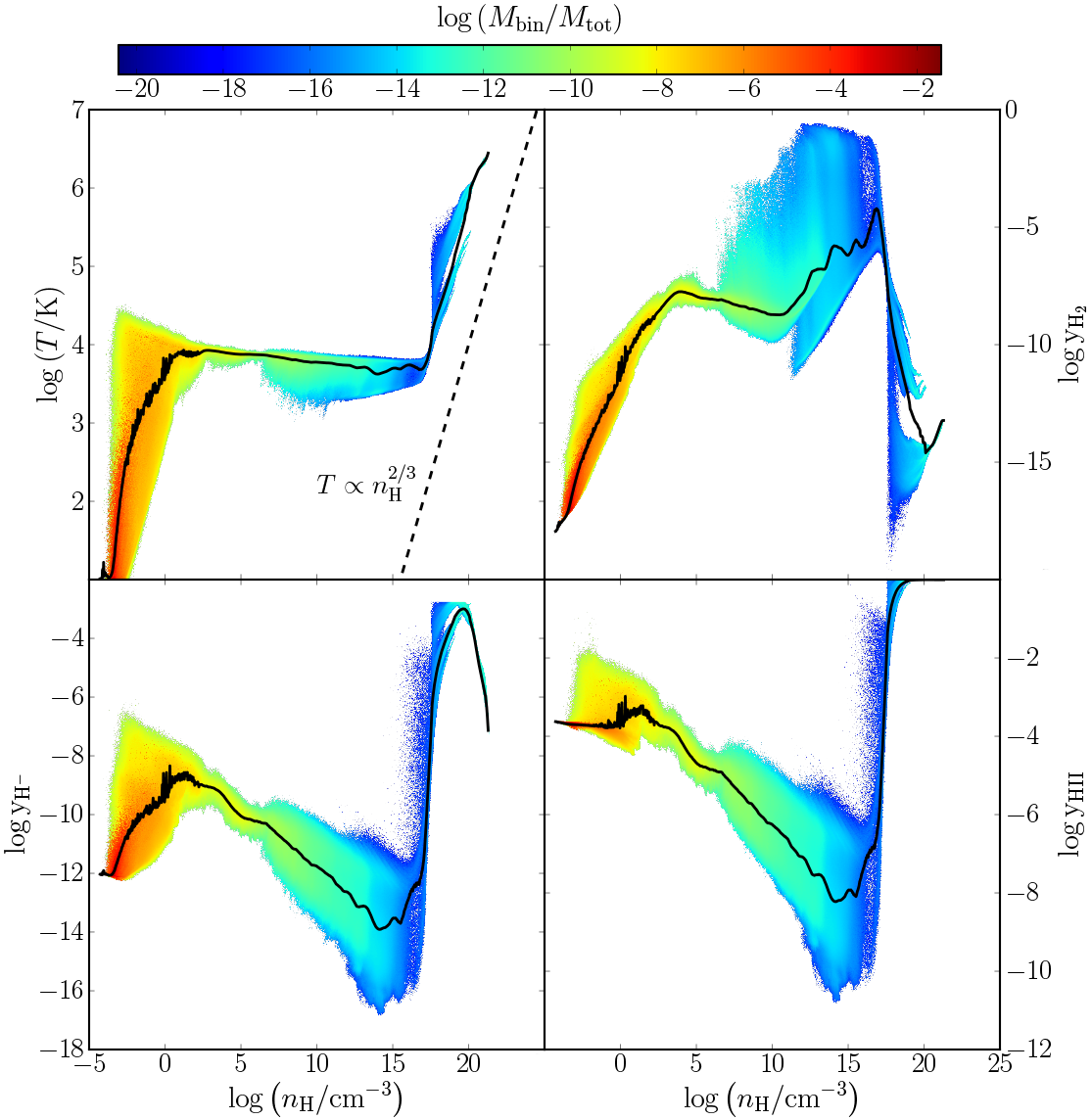}
\caption{Clockwise from the top left panel: distribution of the gas in temperature, $\HH$, \HII, and $\HM$ abundances versus number density of hydrogen nuclei at the end of the simulation. The mass per bin over the total mass in the computational domain is colour-coded from blue (lowest) to red (highest). The solid black lines show the mass-weighted average values. After shock-heating to the virial temperature of $\simeq 10^4\,$K, the gas collapses nearly isothermally to densities of $n \simeq 10^{16}\,\cmmm$. The gas then becomes optically thick to continuum emission and evolves nearly adiabatically. At this point, the \HII abundance dramatically increases from $\sim 10^{-14}$ to unity. The H$_2$ abundance stays below $\simeq 10^{-7}$ due to the LW background, but then increase to $\simeq 10^{-4}$ as three-body reactions set in. However, the H$_2$ abundance never becomes high enough for H$_2$ cooling to become important. The `fingers' visible in the various distributions show the evolutionary paths of individual protostars.}
\label{fig:pspace_T_ab}
\end{center}
\end{figure*}
                
Fig.~\ref{fig:nh_enc_mass_temp_abH2} shows various physical quantities as a function of distance from the densest cell in the halo. The radial profiles are constructed from data of the three resimulations. We proceed by extracting the inner $\simeq 300\,$au from the last resimulation, while the range between $\simeq 300$ and $\simeq 10^5\,$au is taken from the second resimulation. To complete the profiles, the outer region corresponds to data from the first resimulation up to $\simeq 10^{10}\,$au. Due to the self-similarity of the collapse, moving from large to small radii is equivalent to moving from early to late times. Properties plotted in the figure are the number density of hydrogen nuclei, enclosed gas mass, temperature, $\HH$ abundance, $\HM$ abundance, and \HII abundance. These profiles have been calculated using mass-weighted averages of the cells contributing to the radial bins. Colours and line styles represent different evolutionary stages of the gas cloud as described in the legend and the caption of the figure.

\begin{figure*}
\begin{center}
\includegraphics[scale=0.53]{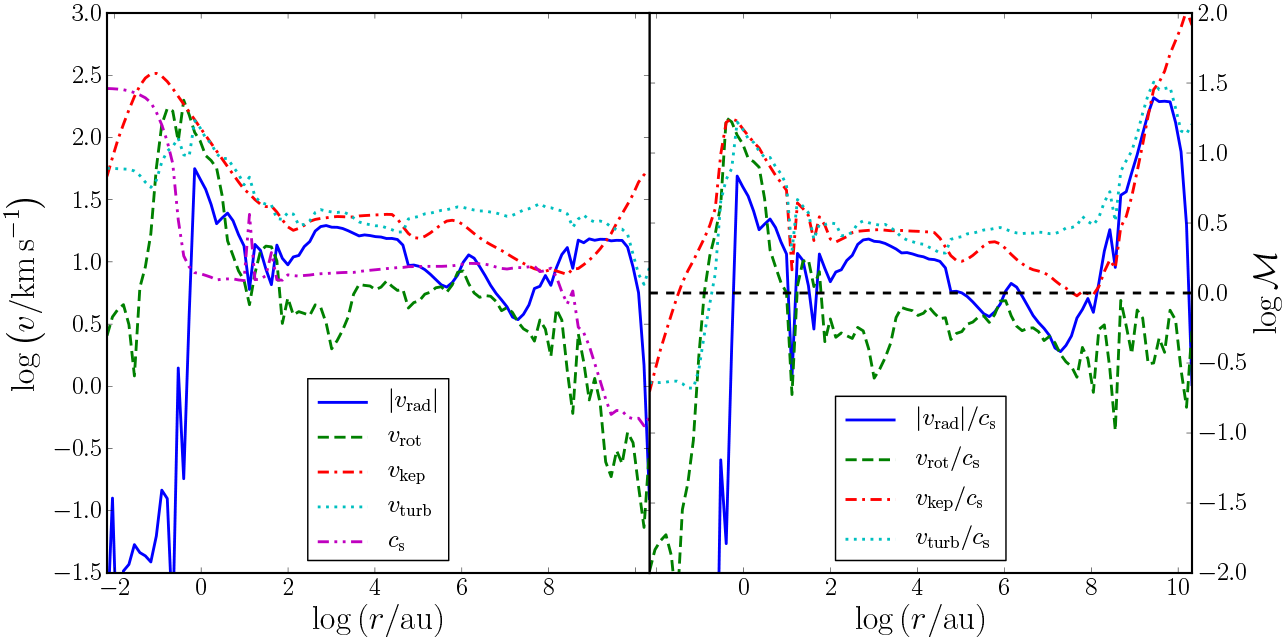}
\caption{Mass-weighted average radial velocity, rotational velocity, Keplerian velocity, turbulent velocity, and sound speed versus radius (left-hand panel), as well as the corresponding Mach numbers (right-hand panel). The turbulent velocity is supersonic with a Mach number of $\simeq 3$ throughout the collapse, while the radial velocity briefly becomes subsonic. Once the disc forms, the rotational velocity becomes nearly equal to the Keplerian velocity, and comparable to the turbulent velocity.}
\label{fig:velocities}
\end{center}
\end{figure*}

\begin{figure*}
\begin{center}
\includegraphics[scale=0.75]{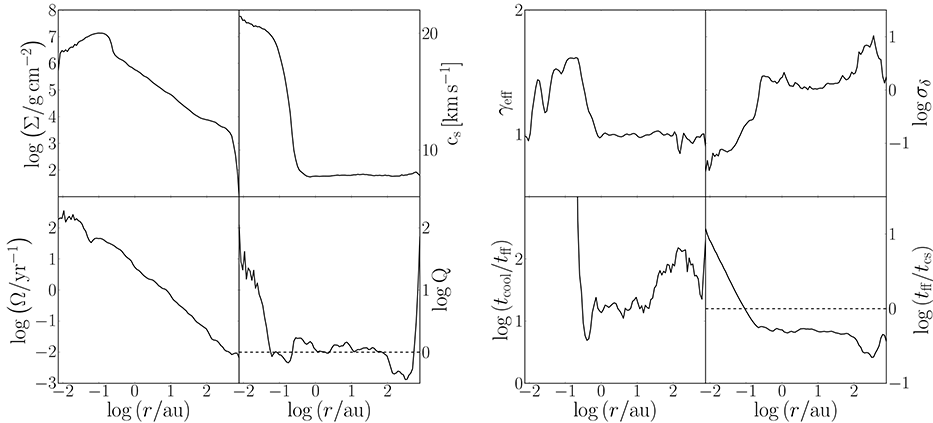}
\caption{Left: from top left to bottom right, the panels show the mass-weighted average surface density, sound speed, orbital frequency, and Toomre parameter versus radius just before the disc fragments. Above $\simeq 0.1\,$au, the power-law profiles of the surface density and rotation speed yield a Toomre parameter that is close to unity, which indicates that perturbations in the disc can grow. Right: effective equation of state, root-mean-squared density contrast, cooling time over free-fall time, and free-fall time over sound-crossing time. The isothermal collapse of the gas on scales $\ga 1\,$au results in $\geff \simeq 1$, while the increasing optical depth of the gas to continuum emission on smaller scales results in an exponent that is closer to that of an adiabatic gas. The cooling time over the free fall time has a local minimum on a scale of $\simeq 1\,$au: this is approximately the radius at which the first fragment forms. The density contrast created by the supersonic turbulence is between $\simeq 1$ and $\simeq 10$. The free-fall time exceeds the sound crossing time on a scale of $\simeq 0.1\,$au, which shows the size of the central, Jeans-unstable clump.}
\label{fig:disc_stability}
\end{center}
\end{figure*}

As the gas collapses into the DM halo, it is shock-heated to the virial temperature. In the central parts of the halo, Ly$\alpha$ cooling becomes important and keeps the gas nearly isothermal at $\simeq 10^4\,$K \citep{Wise_2007}. During this period, the $\HH$ abundance builds up from $\simeq 10^{-16}$ to $\simeq 10^{-8}$, with small spikes due to the existence of shocks in the outer regions of the halo. The \HII abundance increases by about one order of magnitude. The collapse then approximately follows the Larson-Penston solution for an isothermal, self-gravitating gas cloud \citep{Larson_69, Penston_69}, which is described by a density profile following $\rho \propto r^{-2}$. The strong LW background suppresses the formation of $\HH$ and maintains an abundance of $\simeq 10^{-8}$ down to scales of $\simeq 10^3\,$au. This prevents $\HH$ cooling, which in turn leads to a roughly isothermal collapse between $\simeq 10^8$ and $\simeq 1\,$au. Over these scales, the $\HM$ and \HII abundances drop by many orders of magnitude due to recombination. On a scale of $\simeq 10^3\,$au, the $\HH$ fraction increases due to three-body reactions. Up to this point, the radial profiles agree well with those of previous studies \citep{Wise_2008, Latif_2013a, Regan_2014a, Inayoshi_2014}. In the final stage of the collapse, when the primary protostar forms, the gas becomes optically thick to continuum cooling at $\simeq 1\,$au, which results in a rise in temperature of more than two orders of magnitude to $\ga 10^6\,$K. This is accompanied by a drop in the $\HH$ abundance and an increase in both the $\HM$ and \HII abundances. At the end of the simulation, two pronounced spikes in the density profile are clearly visible in the central $\simeq 100\,$au. These correspond to secondary protostars that have formed due to fragmentation in the disc. This will be discussed in detail in Section \ref{subsec:disc}.
        
In Fig.~\ref{fig:pspace_T_ab}, we show the temperature-density distribution of the gas at the end of the simulation, next to those of the $\HH$, $\HM$, and \HII abundances. At low densities, the temperature distribution spans almost six orders of magnitude, reaching as high as $\simeq 10^4~{\rm K}$. A similarly high scatter is present in the $\HH$ and $\HM$ abundances, while the \HII abundance varies only by two orders of magnitude. Up to $\nh\simeq 10^{15}\,\cmmm$, the temperature distribution becomes much narrower, showing the near-isothermal collapse of the gas. Once three-body reactions become important, the distribution of the $\HH$ fraction widens for densities in the range $10^5 - 10^{15}\,\cmmm$, with particles reaching abundances as high as $y_{\rm H_2}\simeq 0.1$. The resulting temperature dispersion leads to an increasing dispersion in the $\HM$ and \HII abundances, while their average values continue to decrease due to recombinations. The values of the $\HH$ abundance are somewhat smaller than those found in \citet{Inayoshi_2014}, but agree with \citet{Latif_2013a}. We therefore do not distinguish between two thermal phases of the gas as in \citet{Inayoshi_2014}. For densities $\ga 10^{18}\,\cmmm$, the formation of the primary and secondary protostars can be recognized as `fingers' of gas in the individual panels, which evolve nearly adiabatically. The high temperatures in the interior of the protostars results in a decrease of the $\HH$ and $\HM$ abundances, and an increase of the \HII abundance to unity.

The radial profiles of the magnitude of the radial velocity, rotational velocity, Keplerian velocity, turbulent velocity, and sound speed at the end of the simulation are shown in the left-hand panel of Fig.~\ref{fig:velocities}. In addition, the right-hand panel shows the Mach numbers of each velocity component. The turbulent Mach number is given by
\begin{equation}
\mathcal{M}_{\rm turb}^2\cs^2 = \sum_{i} {m_i \over M} ({\bf v}_i - {\bf v}^{\rm rad}_i - {\bf v}^{\rm rot}_i)^2,
\label{eq:turb}
\end{equation}
where $\cs$ denotes the sound speed of the radial bin, $M$ the total mass, $i$ the index of a cell contributing to the bin, $m_i$ its mass, ${\bf v}_i$ the velocity, ${\bf v}^{\rm rad}_i $ the radial velocity vector, and ${\bf v}^{\rm rot}_i$ the rotational velocity vector. During the initial free-fall phase, the turbulent component is supersonic with $\mathcal{M}\simeq 3$. In contrast, the Mach number of the rotational velocity remains below unity, indicating the poor rotational support of the cloud at that stage. The trend for each component is roughly maintained once the halo has entered the isothermal collapse phase, with the exception of the Mach number of the radial velocity, which briefly drops to below unity. Down to $\simeq 100\,$au, the rotational velocity oscillates between 0.2 and 0.5 of the Keplerian velocity, indicating a substantial degree of rotational support. It reaches its peak at the edge of the disc on scales $\simeq 1\,$au, where $\vrot\simeq \vkep$. On smaller scales, the primary protostar is characterized by an increase in temperature and thus sound speed, such that all velocity components become subsonic. In addition, $\vrad$ drops precipitously, which shows that the infall rate decreases rapidly within the primary protostar. Similar values for the velocity have been found in previous studies \citep[e.g.][]{Regan_2014a}.

Overall, we find good agreement between our results and previous work. However, some differences exist. The morphology of the halo between $\simeq 10\,{\rm au}$ and $\simeq 10\,{\rm pc}$ is similar to that of \citet{Inayoshi_2014}, but we do not find clumps on larger scales as pointed out by \citet{Regan_2009}, \citet{Latif_2013a}, and \citet{Regan_2014a}. However, this is not surprising, since in our case the gas has not yet had time to settle into a disc on these scales. The radial profiles resemble those of \citet{Latif_2013a} quite well, while \citet{Inayoshi_2014} found a slightly higher $\HH$ abundance, which is also reflected in lower temperatures during the isothermal collapse phase. These differences may be caused by the different chemical networks used in our study (see Section~\ref{subsec:caveats}).

\begin{figure}
\begin{center}
\includegraphics[scale=1.0]{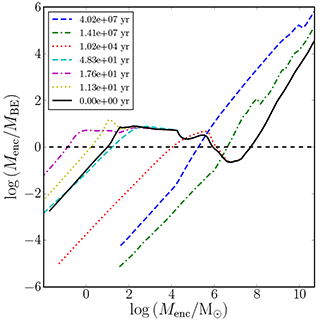}
\caption{Enclosed gas mass over the mass-weighted average BE mass as a function of enclosed gas mass. Colours and line styles are the same as in Fig.~\ref{fig:nh_enc_mass_temp_abH2}. As the halo grows in mass, the BE mass increases due to the rise in the virial temperature, which reduces $\menc /\mbe$. Once the atomic cooling halo is assembled, this ratio exceeds unity on a scale of $\simeq 10^8\,\msun$. Following the onset of runaway cooling due to Ly$\alpha$ emission, the central $10^6\,\msun$ become Jeans-unstable (red dotted line). The minimum Jeans mass of the cloud is indicated by the purple dash-dotted line as $\simeq 0.1~\msun$, which coincides with the initial mass of the primary protostar.}
\label{fig:mbe}
\end{center}
\end{figure}

\subsection{Disc formation and fragmentation}
\label{subsec:disc}

\begin{figure*}
\begin{center}
\includegraphics[scale=0.84]{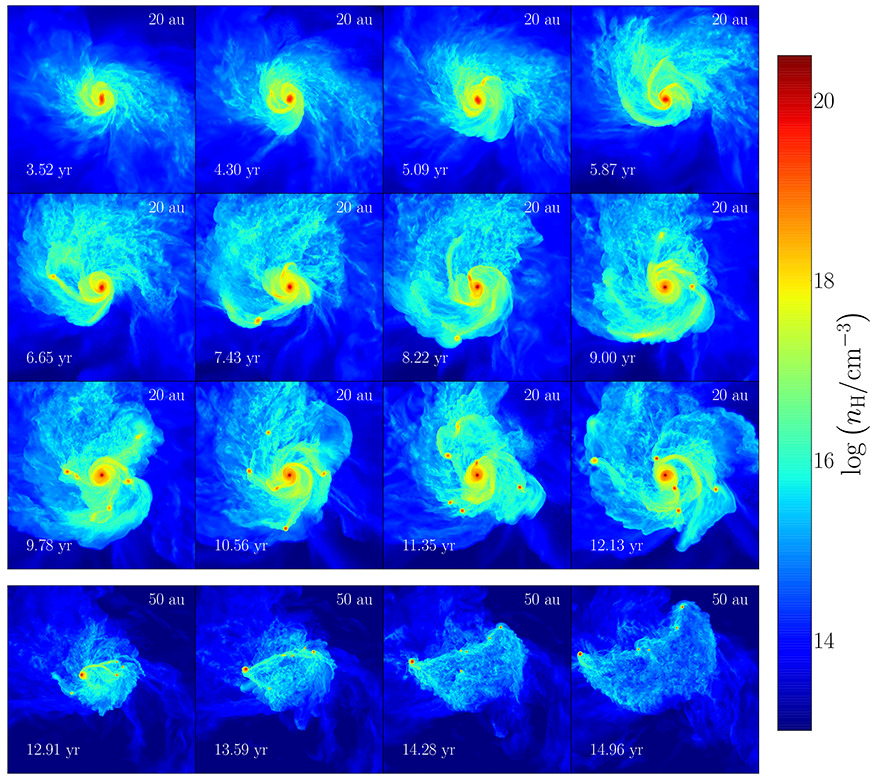}
\caption{Evolution of the protostellar system that forms at the centre of the atomic cooling halo. The number density of hydrogen nuclei is weighted with the square of the density along the line of sight, which is perpendicular to the plane of the disc. The top three rows show cubes with a side length of $20\,$au, centred on the position of the first protostar. The bottom row shows the later evolution of the protostellar system on a somewhat larger scale of $50\,$au, where the centre has been fixed on the position of the primary protostar after $\simeq 12\,$yr. The time is measured from the instant when the density first exceeds $10^{19}\,\cmmm$. The formation of a Keplerian disc around the primary protostar is clearly visible. Shortly thereafter, the disc becomes Toomre-unstable and spiral arms form that transport mass inwards and angular momentum outwards. After $\simeq 6\,$yr, the disc becomes gravitationally unstable and fragments due to the high mass accretion rate from the surrounding cloud on to the disc, and the efficient cooling of the disc by continuum emission. Over the next $\simeq 7\,$yr, an additional five protostars form before three-body interactions lead to the temporary ejection of the primary protostar from the cloud, which disrupts the disc.}
\label{fig:disc}
\end{center}
\end{figure*}    

After the formation of the first protostar, the gas becomes fully rotationally supported in a Keplerian disc. We study its stability by computing Toomre's parameter \citep{Toomre_64}:
\begin{equation}
Q = \frac{\cs \kappa}{\pi G \Sigma},
\label{eq:Q}
\end{equation}
where $\cs$ is the sound speed of the gas, $\kappa$ the epicyclic frequency of the disc, $G$ the gravitational constant, and $\Sigma$ the surface density of the gas. For the case of a Keplerian disc, the epicyclic frequency may be replaced by the orbital frequency $\Omega$. The $Q$ parameter was originally proposed to determine whether perturbations can grow in an infinitely thin, isothermal disc. Later studies have extended this criterion for thick discs, finding that it only deviates by a factor of order unity from the above equation \citep{Wang_2010}. For values greater than $Q_{\rm crit}=1$, the system is stable due to gas pressure and shear by the differential rotation of the disc, while for lower values the system is unstable and hence susceptible to the growth of perturbations. These lead to the formation of spiral arms that transport mass inwards and angular momentum outwards.

\begin{figure*}
\begin{center}
\includegraphics[scale=1.0]{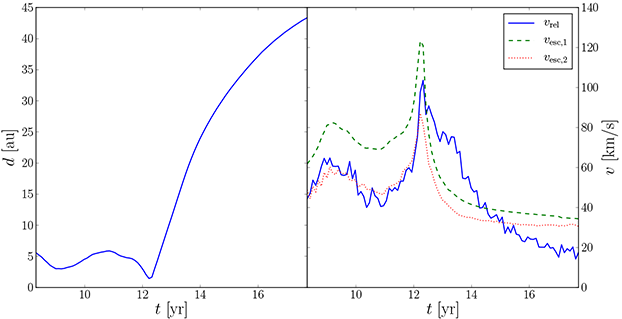}
\caption{Relative distance (left-hand panel) and velocity (right-hand panel, blue solid) between the primary and a secondary protostar. The latter initially orbits around the primary protostar at a distance of $\simeq 4\,$au, but strong gravitational forces due to three-body interactions temporarily eject both protostars from the centre of the cloud after $\simeq 12\,$yr. The relative velocity reaches a peak value of about $\simeq 100\,\kms$, which declines to $20\,\kms$ towards the end of the simulation. For comparison, we also show the escape velocities of both protostars. The green dashed line corresponds to the primary protostar, and the red dotted line to the secondary protostar. Since the relative velocity decreases to well below the escape velocity, both protostars will likely return to the centre of the cloud.}
\label{fig:dist_and_vel}
\end{center}
\end{figure*}

Radial profiles for the gas surface density, sound speed, orbital frequency, and Toomre parameter are shown in the left-hand panel of Fig.~\ref{fig:disc_stability}. We compute the profiles using mass-weighted spherical shells centred on the densest cell in the halo of the final resimulation. The surface density increases as $\Sigma \propto r$ in the interior of the protostar, where the density is almost constant. On larger scales, the radial dependence changes to $\Sigma \propto r^{-1}$, as deduced from the relation $\rho \propto r^{-2}$ for isothermal collapse, while the orbital frequency roughly follows $\Omega \propto r^{-1}$. On scales between 0.1 and $100\,$au, the radial dependence of $\Sigma$ and $\Omega$ thus cancel each other, such that $Q$ remains roughly constant around unity. In the interior of the protostar, $Q$ increases due to the increase in the sound speed and the different radial scaling between $\Sigma$ and $\Omega$. Since the value of $Q$ is roughly equal to the critical value, the disc is prone to perturbation growth.

Further properties at the time when the primary protostar has just formed and is surrounded by a disc that has not yet fragmented are shown in the right-hand panel of Fig.~\ref{fig:disc_stability}. From the top left to the bottom right, the panels show the effective equation of state, root-mean-squared density contrast, cooling time over free-fall time, and free-fall time over sound-crossing time. In the outer region of the disc, on scales $\ga 1\,$au, the equation of state is characterized by $\geff \simeq 1$, as expected for isothermal collapse. The density contrast is roughly constant around unity, while the interior of the primary protostar is characterized by values close to 0.1. Here, $\geff$ increases to $\simeq 1.2 - 1.5$ as a reflection of the temperature increase by almost two orders of magnitude in the central $\simeq 1\,$au. The cooling time remains well above the free-fall time in the inner $100\,$au of the halo and down to $\simeq 1\,$au, the scale at which the gas becomes optically thick to continuum cooling. The free-fall time remains below the sound-crossing time down to $\simeq 0.1\,$au, showing the gravitational instability of the cloud down to this scale.

\subsection{Minimum fragment mass}
\label{subsec:mbe}

Further evidence for the gravitational instability of the gas is presented in Fig.~\ref{fig:mbe}, where we plot the enclosed gas mass over the locally estimated Bonnor-Ebert \citep[BE;][]{Ebert_1955, Bonnor_1956} mass as a function of enclosed gas mass. Colours and line styles are the same as in Fig.~\ref{fig:nh_enc_mass_temp_abH2}. The profiles have been computed using spherical shells centred on the densest cell, where the BE mass is calculated as the mass-weighted average among cells within a given radius according to:
\begin{equation}
\mbe \simeq 15\,\msun\left( {\nh \over \cmmm} \right)^{-1/2}\left({ T \over {\rm K}}\right)^{3/2} \mu^{-3/2}\gamma^2.
\label{eq:mbe}
\end{equation}
During the initial collapse, the ratio of enclosed gas mass to BE mass decreases as a consequence of the rise in temperature as the gas is shock-heated. The enclosed gas mass surpasses $\mbe$ at $\menc \simeq 10^8\,\msun$, which is in agreement with the mass of the halo. As the halo keeps accreting, another region where the ratio exceeds unity emerges at about $\simeq 10^6\,\msun$. This marks the initial Jeans instability of the cloud. From $\simeq 10^6\,\msun$, the point where $\menc$ surpasses $\mbe$ moves down to $\simeq 0.1\,\msun$ when the densest cell first reaches $10^{19}\,\cmmm$. This is the minimum fragment mass and coincides with the initial mass of the protostar formed at the centre of the halo. From then on the temperature of the central object increases, which is translated into an increase of the BE mass, and hence a decrease of the $\menc/\mbe$ ratio. As a result, the point at which this ratio equals unity briefly moves up to $\simeq 10\,\msun$ and always stays above $\simeq 1\,\msun$.

\subsection{Protostellar system}
\label{subsec:protosystem}

The fragmentation of the disc into a small protostellar system is shown in Fig.~\ref{fig:disc}. The top three rows show cubes of side length $20\,$au centred on the position of the primary protostar, while the cubes of the last row are $50\,$au wide with the centre fixed on the position of the primary protostar after $\simeq 12\,$yr. In total, we present 16 different output times, which are measured with respect to the point in time at which the densest cell first exceeds $10^{19}\,\cmmm$. During the first $\simeq 6$ yr, perturbations grow between $1$ and $10\,$au in the form of spiral arms. After $\simeq 7\,$yr, they become gravitationally unstable and the first secondary protostar forms. In the next $\simeq 2\,$yr, the efficient cooling of the gas results in the formation of additional protostars, and after $\simeq 12~{\rm yr}$ a small protostellar system with six members has emerged. Shortly thereafter, three-body interactions and strong tidal forces during a close passage of a secondary protostar and the primary results in the disruption of the disc. Both protostars are ejected from the centre of the cloud. The sequence in the bottom row of Fig.~\ref{fig:disc} shows the evolution of this interaction and how both protostars move away from each other.

\begin{figure}
\begin{center}
\includegraphics[scale=0.86]{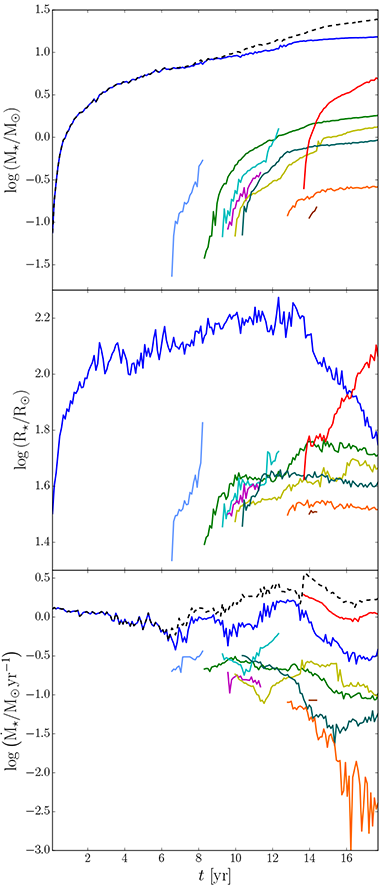}
\caption{Stellar mass, radius, and accretion rate of all protostars formed in the simulation. Each line corresponds to an individual protostar, and the black dashed lines shows the total mass and accretion rate, respectively. Initially, the mass budget is entirely dominated by the primary protostar (blue line), which grows from $\simeq 0.1$ to $\simeq 15\,\msun$ at a rate of $\simeq 1\,\msunyr$, while its radius swells to well over $100\,$au. Once the primary protostar is expelled from the centre, its accretion rate and size drop significantly. The protostar formed in the secondary clump (red line) grows to about $5\,\msun$ at a rate of $\ga 1\,\msunyr$. The other protostars stay below $\simeq 2\,\msun$, and thus do not contribute significantly to the total protostellar mass.}
\label{fig:protoproperties}
\end{center}
\end{figure}

To quantify the interaction between both protostars, Fig.~\ref{fig:dist_and_vel} shows the relative distance and velocity between the protostars over time. For comparison, we also include the escape velocity of both protostars, using the enclosed mass in a spherical region around their respective centres, with radii equal to their separation. Once the secondary protostar forms, it orbits at a roughly constant distance of $\simeq 4\,$au from the primary protostar, but they soon move together and their separation decreases to $\simeq 1\,$au. Shortly thereafter, three-body interactions eject both protostars from the centre of the halo. This is reflected by a high relative velocity with a peak value of $\simeq 100\,\kms$, which is followed by a gradual drop in the relative velocity. The parabolic shape of the relative distance suggests that it may reach a point of turnaround and the protostars will begin to re-collapse towards the centre. This trend is supported by the declining profile of the relative velocity and the fact that it has fallen well below the escape velocity by the end of the simulation.

In Fig.~\ref{fig:protoproperties}, we show the mass, radius, and accretion rate of all protostars over time. The solid lines correspond to individual protostars, while the black dashed lines denote the total mass and accretion rate, respectively. The radius of a protostar is calculated as the distance at which the Rosseland mean opacity reaches its maximum value \citep{Stacy_2013}. The protostellar mass is given by the mass enclosed within that radius, and the accretion rate by the time derivative of the enclosed mass. A total of eight secondary protostars form during the evolution and fragmentation of the disc. Out of these, four survive until the end of the simulation. The rest merge with other protostars or are tidally disrupted. During the first $\simeq 6\,$yr after the formation of the primary protostar, its mass builds up from $\simeq 0.1$ to $\simeq 6.4\,\msun$ at a rate of roughly $1\,\msunyr$, in agreement with previous work \citep{Latif_2013a, Inayoshi_2014}. Its radius increases from $\simeq 32$ to $\simeq 136\,\rsun$. The second protostar forms after $\simeq 7\,$yr with an initial mass of $\simeq 0.02\,\msun$ and a radius of $\simeq 22\,\rsun$. Most of the gas is accreted by the primary protostar, while the second protostar only accretes at a rate of $\simeq 0.3\,\msunyr$ before it is tidally disrupted.

\begin{figure*}
\begin{center}
\includegraphics[scale=0.82]{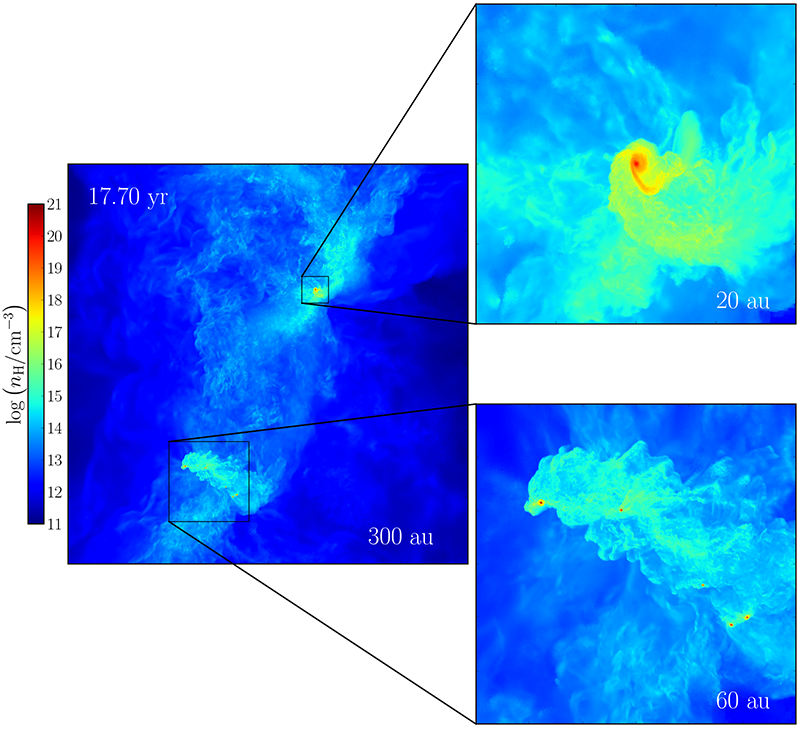}
\caption{Simultaneous collapse of a secondary gas clump at a distance of $150\,$au from the primary clump in the atomic cooling halo. The number density of hydrogen nuclei is weighted with the square of the density along the line of sight in a cube with a side length of $300\,$au. The panels on the right show zoom-ins on the primary and secondary clumps with a side length of 60 and $20\,$au, respectively. While strong interactions occur in the central protostellar system, a second clump has collapsed and is in the early stages of its evolution. Ultimately, the clumps may evolve into a wide binary system.}
\label{fig:box}
\end{center}
\end{figure*}

Shortly thereafter, the disc fragments vigorously and gives rise to a protostellar system characterized by a massive primary protostar with $\simeq 9.2\,\msun$ and a radius of $\simeq 160\,\rsun$, while the secondary protostars only have masses between $\simeq 0.06$ and $\simeq 0.6\,\msun$, and radii in the range $\simeq 26 - 41\,\rsun$. The accretion on to the secondary protostars results in a slight decrease of the accretion rate on to the primary protostar of $\simeq 0.5\,\msunyr$, while the total accretion rate remains roughly constant at $1.5\,\msunyr$. After $\simeq 13~{\rm yr}$, the primary protostar is expelled from the centre of the halo and the disc is disrupted (see bottom row of Fig.~\ref{fig:disc}). As the primary protostar is deprived of gas, its accretion rate drops to $\la 0.3\,\msunyr$, and its radius decreases from $\simeq 160$ to $\simeq 56\,\rsun$. Its final mass is $\simeq 15\,\msun$.

The formation of a protostellar system has recently been reported in studies of minihaloes that cool via $\HH$ lines \citep{Clark_2008, Clark_2011, Greif_2011}. Initial masses of protostars in atomic cooling haloes are an order of magnitude higher, while the accretion rates exceed those in minihaloes by about three orders of magnitude. Other studies have found similar values for the initial protostellar masses and accretion rates \citep{Regan_2009, Latif_2013a, Regan_2014a, Inayoshi_2014}.
        
\subsection{Secondary clump}
\label{subsec:secondary}

About $13\,$yr into the evolution of the protostellar system, a second clump collapses at a distance of about $150\,$au from the primary clump. Fig.~\ref{fig:box} shows both clumps in a cube with a side length of $300\,$au at the end of the simulation, with smaller cubes showing zoom-ins on to the individual clumps. The zoom-in of the secondary clump shows a protostar with a disc and spiral arms similar to the early evolutionary stages of the primary clump. The protostar in the secondary clump is denoted by the red line in Fig.~\ref{fig:protoproperties}. Its mass quickly grows from $\simeq 0.2$ to $\simeq 4.9\,\msun$, and its radius increases from $\simeq 42$ to $\simeq 116\,\rsun$. Despite its later formation, it accretes more rapidly than the first protostar. Ultimately, both clumps may evolve into a wide binary system.

\subsection{Caveats}
\label{subsec:caveats}

Previous studies that investigated the collapse and fragmentation of gas in atomic cooling haloes did not have sufficiently high resolution to self-consistently follow the formation of protostars at the centre of the cloud. We have attempted to address this shortcoming by performing a simulation that is not resolution-limited. Nevertheless, we have neglected some physical processes that might affect the fragmentation of the cloud. In particular, we have assumed that the optically thin regime for atomic hydrogen cooling extends up to densities $\simeq 10^{16}\,\cmmm$. In reality, the gas becomes optically thick to Ly$\alpha$ radiation at densities of $\simeq 10^6\,\cmmm$, and then free-bound continuum emission of $\HM$ becomes the main cooling agent. Previous studies have found that this kind of cooling may lower the temperature by up to a factor of $2$ in the range $n \simeq 10^{15} - 10^{20}\,\cmmm$ compared to our study \citep{Omukai_2001, Inayoshi_2014}. A lower temperature should translate into a lower Toomre parameter, which would enhance the fragmentation seen in our simulation. In addition, at $\nh \ga 10^{16}\,\cmmm$ we have introduced an artificial cut-off for continuum cooling in order to approximately reproduce the density-temperature relation found in \citet{Omukai_2001}. This simplification may also affect the thermal and gravitational stability of the gas.

Another factor that might influence the temperature of the disc is the heating from the accretion luminosity of the primary protostar. The accretion luminosity is given by
\begin{equation}
\Gamma_{\rm acc}=\kappa_{\rm P} \left( {L_{\rm acc} \over 4\pi r^2} \right),
\end{equation}
where $\kappa_{\rm P}$ is the Planck mean opacity, $r$ the distance from the source, and $L_{\rm acc} = G\mstar \mstardot / \rstar$ the accretion luminosity. The effects of the accretion luminosity have been discussed in similar studies that focused on minihaloes \citep{Greif_2011, Smith_2011}. They found that the additional heating of the gas may slightly delay fragmentation, but does not prevent it. The photospheric temperature of the protostar of $\simeq 8000\,$K during the early stages of the collapse is too low to produce significant amounts of ionizing radiation. \citet{Latif_2013a} investigated the influence of accretion luminosity in atomic cooling haloes. Assuming a power-law relation between the mass and the radius of the star, and an accretion rate of $\simeq 1\,\msun\,{\rm yr}^{-1}$, they computed an accretion luminosity of $\simeq 2\times 10^{-4}\,{\rm erg}\,{\rm cm}^{-3}\,{\rm s}^{-1}$ for a $500\,\msun$ clump with a size of $\simeq 100\,$au and a temperature of $\simeq 8000\,{\rm K}$. This value is comparable to the energy emitted by Ly$\alpha$ cooling, and may exceed it once the mass of the clump reaches $\simeq 1000\,\msun$. However, \citet{Latif_2013a} found that this difference only translates into an increase of the temperature by $\simeq 500\,{\rm K}$. Since we investigate the evolution of the protostellar system at even earlier times, when the mass of the protostar is much lower, the effects of the accretion luminosity are expected to be even smaller.

Next to the aforementioned cooling and heating processes, we do not include the effects of magnetic fields. These are expected to become dynamically important in minihaloes as well as atomic cooling haloes \citep[e.g.][]{Xu_2008, Schleicher_2010, Sur_2010, Peters_2012, Peters_2014, Schober_2012, Turk_2012, Latif_2013c}. Indeed, \citet{Latif_2014b} found that the magnetic pressure provides additional support against gravity and delays or suppresses fragmentation. Future simulations should therefore include magnetic fields as well as a more detailed chemical and thermal model.

\section{Summary and Conclusions}
\label{sec:conclusion}

We have performed the highest-resolution cosmological simulation to date of the formation and evolution of a protostellar system in an atomic cooling halo. We follow the collapse of the gas from a few Mpc down to $\la 0.01\,$au, spanning almost 13 orders of magnitude in scale, and reaching densities as high as $\nh\simeq 10^{22}\,\cmmm$. The simulation includes an equilibrium/non-equilibrium primordial chemistry solver that evolves five species (H, $\HH$, $\HM$, $\HP$, and $\e$), and includes $\HH$ line emission, $\HH$ collision-induced emission, Ly$\alpha$ cooling, and inverse Compton cooling. Additionally, we have included a uniform LW background of strength $J_{21}=10^5$ to prevent star formation in progenitor haloes.

During the initial collapse, the gas is shock-heated to the virial temperature of about $10^4~{\rm K}$. The molecular hydrogen abundance briefly increases due to the presence of supersonic shocks, but the external radiation background photodissociates H$_2$ to a level of $y_{\rm H_2}\sim 10^{-7}$ within the halo. As a result, runaway collapse due to Ly$\alpha$ cooling ensues once the virial mass has risen to $\simeq 5\times 10^7\,\msun$. The central gas cloud becomes Jeans-unstable with a mass of $\simeq 10^6\,\msun$ and collapses nearly isothermally over many orders of magnitude in density, characterized by a profile of the form $\rho \propto r^{-2}$. At densities $\nh\sim 10^6\,\cmmm$, the gas becomes optically thick to Ly$\alpha$ emission and effectively cools via free-bound continuum emission of $\HM$ up to a density of $\nh\sim 10^{16}\,\cmmm$, where the continuum emission is trapped. The average $\HH$ abundance increases to $y_{\rm H_2}\sim 10^{-4}$ at $\nh\ga 10^{10}\,\cmmm$ due to three-body reactions, but never becomes high enough for H$_2$ line emission to become important. The $\HP$ abundance declines to $\simeq 10^{-8}$ due to recombinations before increasing to unity for densities $\ga 10^{16}\,\cmmm$, where the gas evolves nearly adiabatically and a protostar with an initial mass of $\simeq 0.1\,\msun$ is formed.

Following the formation of the primary protostar, the gas settles into a Keplerian disc. The Toomre parameter within the disc is close to unity, such that perturbations can grow. The emerging spiral arms feed gas on to the primary protostar at a rate of $\simeq 1\,\msunyr$. However, this is not sufficient to process the mass that accretes from the surrounding cloud on to the disc. In combinations with the efficient cooling of the gas via continuum emission, the disc becomes gravitationally unstable and a secondary protostar forms after only $\simeq 7\,$yr. The disc continues to fragment, such that after $\simeq 18\,$yr a total of eight secondary protostars have formed. By the end of the simulation, four of these have survived, while the rest have merged away or are tidally disrupted. The primary protostar has grown to a mass of $\simeq 15\,\msun$, while all other secondary protostars have masses $\la 2\,\msun$. Three-body interactions lead to the temporary ejection of the primary protostar from the disc after $\simeq 12\,$yr, which is disrupted in the process. However, an analysis of the relative velocity of the protostars shows that it is well below the escape velocity. It will therefore likely return to the centre of the cloud. After $\simeq 13\,{\rm yr}$, a second clumps collapses at a distance of $\simeq 150\,$au from the primary clump. It has not yet fragmented and contains a single protostar that rapidly grows to $\simeq 5\,\msun$. If this clump show a similar pattern of rapid migration and merging, the cloud may evolve into a wide binary system.

Despite the temporary ejection of the primary protostar from the centre of the cloud, subfragmentation likely does not substantially impede its growth. Once it returns to the centre of the cloud, its accretion rate will likely again increase to $\simeq 1\,\msunyr$. In addition, the secondary protostars formed in the disc quickly migrate to the centre of the cloud, where they merge with the primary protostar. They are also typically $10$ times less massive than the primary protostar, which has accreted $\simeq 15\,\msun$ by the end of the simulation, while the most massive secondary protostar has only grown to $\simeq 1.5\,\msun$. Most of the accreted material thus does not stem from other protostars, but from the bar-like instabilities in the disc. The secondary clump may be a much more potent candidate for accreting mass that may have otherwise been accreted by the primary clump, but even in this case the growth of the most massive protostar would be reduced by at most a factor of $2$. It thus appears that fragmentation is not a significant barrier for forming at least one massive BH seed per atomic cooling halo, assuming that the LW background is high enough to prevent H$_2$ cooling. Recent simulations have shown that this may indeed be the much more limiting factor \citep{Latif_2014a, Latif_2014c, Regan_2014b}.

One of the main caveats of this study is the simplified chemistry and cooling network. Future work should include a more detailed chemical model, such as that used in \citet{Inayoshi_2014}. It may also become possible to treat the radiative transfer of the various line and continuum processes \citep[e.g.][]{Greif_2014}. Finally, the influence of magnetic fields may be investigated with modules that have already been implemented in {\sc arepo} \citep{Pakmor_2011}. The influence of the radiation may not be that strong, since it is difficult to heat the gas above $\simeq 10^4\,$K, while magnetic fields may have a substantial effect on the thermal and gravitational stability of the cloud \citep[e.g.][]{Latif_2013c, Latif_2014b}. The additional support provided by magnetic fields may reduce the ability of the gas to fragment, and further increase the accretion rate of the primary protostar.

\section*{Acknowledgements}

FB would like to thank Simon Glover, Paul Clark, John Regan, Muhammad Latif, and Kohei Inayoshi for stimulating discussions and feedback during the conference `The physics of first stars and galaxy formation' at The Higgs Centre for Theoretical Physics of the University of Edinburgh. VS acknowledges support by the European Research Council under ERC-StG EXAGAL-308037. The simulations were carried out at the Texas Advanced Computing Center (TACC) under XSEDE allocation AST130020.




\end{document}